\def\fg1{ \def\do{{$\!$.}}
\begin{figure}[bht] \unitlength1cm \begin{picture}(14,3.2)
\put(.85,.4){$\d\P$} \put(2.4,.4){$=$}
\put(7,.5){$+$}  \put(12.7,.4){$-$}  \put(13.8,.4){$\P^{[-]}$}
\thicklines \put(5,.5){\circle{.4}}  \put(5,.5){\circle*{.09}}
            \put(9,.5){\circle{.4}}  \put(9,.5){\circle*{.09}}
            \put(11,.5){\circle{.4}} \put(11,.5){\circle*{.09}}
\thinlines  \put(4.5,.4){\line(1,0){.32}}
\put(5.18,.4){\line(1,0){.32}} \put(8.5,.5){\line(1,0){.3}}
\put(11.2,.5){\line(1,0){.3}} \put(4.82,1.4){\oval(1.44,1.6)[bl]}
\put(5.18,1.4){\oval(1.44,1.6)[br]} \put(5,1.3){\oval(1.8,1.8)[t]}
\put(10,.7){\oval(2,1.1)[t]}     \put(10,.3){\oval(2,1.1)[b]}
\put(5,2.2){\circle*{.25}}
\put(10,1.25){\circle*{.25}}  \put(10,-.25){\circle*{.25}}
\end{picture} \caption[f1]{\label{f1}{\frm
   The actual next-to-leading order contributions to the
   polarization function. The gluon propagators are resummed
   (as indicated by a black bullet), and each vertex $\bigodot$
   is made up of a bare piece (dot) and a hard thermal loop.
}}  \vspace{.2cm}   \end{figure}  }
\def\ov#1{ \overline{#1} }   \def\cl#1{{\cal #1}}
\def\({\left(}        \def\){\right)}
\def\lk{\,\left[ \,}  \def\rk{\,\right] \,}
      
    \def\Tr{ \, {\rm Tr} \, }
\def\lsim{\,\lower 4pt\hbox{$\sim\!\!\!\!\!$}\raise
      2pt\hbox{$<$}\,}
\def\gsim{\,\lower 4pt\hbox{$\sim\!\!\!\!$}\raise
      2pt\hbox{$>$}\,}
\def\mn{ _{\mu \nu} }        \def\omn{ ^{\mu \nu} }
\def\be{ \begin{equation} }  \def\bea{ \begin{eqnarray} }
\def\ee{ \end{equation} }    \def\eea{ \end{eqnarray} }
\def\nonu{ \nonumber }       
\def\P{ {\mit\Pi} }          \def\X{ {\mit\Xi} }
\def\0{\over } \def\6{\partial }
\let\a=\alpha   \let\d=\delta  \let\e=\varepsilon
   \let\o=\omega    
  \let\D=\Delta  \let\G=\Gamma  \let\O=\Omega
   \def\wu#1{\sqrt{{#1} \,}^{ \hbox to0.2pt{\hss$
       \vrule height 2pt width 0.6pt depth 0pt $} \;\! } }
   \def\pfeil{_\rightharpoonup}  \def\leer{\phantom{a}}
   \def\opf{\buildrel \pfeil \over \leer}
   \def\jvv{j \lower0.4pt\hbox to 2pt{\hss $\opf$}}
   \def\jv{j \lower0.2pt\hbox to 1.4pt{\hss $\opf$}}
   \def\ivv{i \lower0.4pt\hbox to 2pt{\hss $\opf$}}
   \def\iv{i \lower0.2pt\hbox to 1.4pt{\hss $\opf$}}
   \def\hq{h \raise0.2pt\hbox to 0.4pt{\hss $^-$}}
   \def\vk#1{\hbox{$\buildrel           \pfeil \over #1$}}
   \def\vkk#1{\hbox{$\buildrel   \;     \pfeil \over #1$}}
   \def\vkkk#1{\hbox{$\buildrel  \, \;  \pfeil \over #1$}}
   \def\grpf{\displaystyle  _\rightharpoonup}
   \def\vg#1{\hbox{$\buildrel       \grpf \over #1$}}
   \def\vgg#1{\hbox{$\buildrel  \;  \grpf \over #1$}}
\def\fzz{f} \def\bzz{b} \def\dzz{d} \def\gzz{g} \def\hzz{h}
\def\jzz{j} \def\kzz{k} \def\lzz{l} \def\mzz{m} \def\wzz{w}
\def\tzz{t} \def\izz{i} \def\bezz{\beta} \def\dezz{\delta}
\def\xizz{\xi} \def\pszz{\psi} \def\vthzz{\vartheta}
\def\uph{ \! \mathop{\vphantom{a}} } \def\dph{ \vphantom{a} }
\def\vc#1{\def\tast{\noexpand#1} \def\test{#1}
\ifcat\tast\bzz
\ifx\test\fzz \vkkk f \uph \else   \ifx\test\bzz \vkk b \uph \else
\ifx\test\dzz \vkkk d \uph \else   \ifx\test\gzz \vkk g \dph \else
\ifx\test\hzz \vkk h \uph \else    \ifx\test\izz \ivv \dph \else
\ifx\test\jzz \jvv \dph \else      \ifx\test\kzz \vkk k \uph \else
\ifx\test\lzz \vkk l \uph \else    \ifx\test\tzz \vkk t \uph \else
\ifx\test\mzz \vg m \dph \else     \ifx\test\wzz \vg w \dph \else
\ifnum \lq#1<91 \vgg #1 \uph \else \vk #1 \dph
  \fi \fi \fi \fi \fi \fi \fi \fi \fi \fi \fi \fi \fi   \else
\ifx\test\bezz \vkk \beta \uph \else  \ifx\test\pszz \vkk \psi
\dph \else \ifx\test\dezz \vkk \delta \uph \else \ifx\test\xizz
\vkk \xi \uph \else \ifx\test\vthzz \vkk \vartheta \uph \else
\vk #1 \dph  \fi \fi \fi \fi \fi \fi }
\let\thq=\theequation    \font\frm=cmr10   \def\eq#1{(\ref{#1})}
\def\parag#1{ \vspace{1.5cm} \hspace{.08cm} \parbox{15cm}{{\bf #1}
       \vspace{1.1cm} } \hfill \vphantom{a} \nopagebreak \indent }
\def\W{\hbox{${\cl W}$}} \def\X{\hbox{${\cl X}$}}
\def\Y{\hbox{${\cl Y}$}} \def\Z{\hbox{${\cl Z}$}} \def\V{\cl V}
\def\low{\lower 1pt\hbox{$\vphantom{a}$}}
\def\zzz{\,\underline{\underline{\low z\!}}\;}
\def\JJJ{\,\underline{\underline{\low\hbox{$J$}\!}}\,}
   \def\dit{{\rm ditto}^{-}}    \def\pt#1{{\bf (#1)}$\,$}
   \def\Pc{\mbox{\boldmath$\,\wp\,$}}
   \def\Cc{\hbox{\boldmath$\,\chi\,$}}
   \def\fY{{\,}^*\!\!\;\Y}      \def\fZ{{\,}^*\!\Z}
   \def\fG{{\,}^*\!\!\;\G}      \def\bG{{\,}^\circ\!\!\;\G}
   \def\I{$\cl I \,$}
\documentstyle[12pt,twoside]{article}

\begin{document}
\begin{titlepage}
\begin{flushleft}   DESY 94-240 \\ ITP-UH 20/94 \\
                    hep-ph/9501229  \end{flushleft}
\vspace{-1.8cm}
\begin{flushright}  December 1994 \hspace{.7cm} $ $
                    \end{flushright}   \vfill \vskip 2.5cm
\begin{center}
   {\Large \bf On the Gluon Plasmon Self-Energy at $O(g)$ } \\
   \vskip 1.2cm \vfill
   {\large Fritjof Flechsig \ and \ Hermann Schulz} \\ \bigskip
   \bigskip  {\sl    Institut f\"ur Theoretische Physik,
                     Universit\"at Hannover \\
              Appelstra\ss e 2, D-30167 Hannover, Germany \\  }
   \vskip 2cm \vfill     {\large    ABSTRACT}
\end{center}
\begin{quotation}  \ \
   The next-to-leading order contribution
   $\d\P\omn ( \o , \vc q )$ to the polarization function of
   the hot gluon system is analysed at non-zero wave vectors
   $\vc q \,$. Using Braaten-Pisarski resummation and general
   covariant gauges, $\d\P\omn$ is found to be gauge-fixing
   independent and transverse on the longitudinal mass-shell.
   The real part of the longitudinal component $\d\P_\ell$ is
   UV and IR stable (for real $q$). At imaginary $q$ it is IR
   singular, and at the point $\o=0$, $q^2=-3m^2$ it coincides
   with the result of Rebhan for next-to-leading order Debye
   screening. When $q$ approaches the lightcone, $\d\P_\ell $
   diverges like $1 / \wu {\o^2 - q^2} $, reflecting the
   breakdown of the Braaten-Pisarski decomposition scheme
   in this limit.
\end{quotation}      \vspace{.2cm}
\begin{flushleft} e-mail$\,$:
       flechsig@itp.uni-hannover.de \end{flushleft}
\end{titlepage}
%
%
\let\dq=\thq \renewcommand{\theequation}{1.\dq}
\setcounter{equation}{0}

\parag {1. \ Introduction }

Infinite temperature is the limit in which QCD can be solved.
After the complete ''zeroth approxi\-mation'', or $O(1)$, was
worked out \cite{BP,BPward} and had been cast into the form of an
effective action \cite{eff}, there were several studies of the
Braaten-Pisarski resummed perturbation theory in ''true first
order'', or $O(g)$, whose results are (and must be) automatically
gauge-fixing independent. The gluon plasmon damping rate at zero
wave vector \cite{damp}, the Debye screening length \cite{Reb}
and the first correction to the plasmon frequency at zero wave
vector \cite{nt} (henceforth referred to as ''$\,$\I$\,$'') are
such $O(g)$ phenomena and do indeed exhibit the required
independence. The notorious difficulties, which are caused
by the non-abelian infra-red instabilities (magnetic mass)
\cite{linde,GPY}, might be solved by a second stage of
effective theories \cite{Bra}, thereby justifying all results
whose derivation does not hit this perturbative barrier.
Problems of another type, arising when the plasmon dispersion
line intersects the lightcone, are solved for a toy-model
\cite{KRS} of the gluon plasma, which is hot scalar
electrodynamics.

In this note, we concentrate on the extension to non-zero
wave vector argument of the gluon self-energy (or polarization
function) $\d\P\omn (\o , \vc q )$ at $O(g)$. Special attention
is paid to the real part of its longitudinal component
$\d\P_\ell \equiv \Tr ( B \P ) $ (for $B$ see \S ~2),
which determines the spectrum of the collective mode (plasmon).
For simplicity, quarks are kept out of our hot black body volume.
Hence its Lagrangian is given by
\be \label{1Lag}
 {\cal L} \; = \;
       - {1 \0 4} \, F\mn^{\,\enskip a} F^{\mu \nu \; a}
       - {1 \0 2 \alpha} \,\( \6^\mu \! A_\mu^a \)^2
       + \, \ov{c}^{\, a} \6^\mu D_\mu^{ab} c^b \quad .
\ee
with  $ F\mn^a = \6_\mu A_\nu^a - \6_\nu A_\mu^a
        + gf^{abc} A_\mu^b A_\nu^c $ and
$ D_\mu^{ab} = \d^{ab} \6_\mu - g f^{abc} A_\mu^c\,$.
The diagrams relevant to the order $O(g)$ under study are
shown in fig.~1. As the present work is intimately related
to the earlier paper \I, we refer to it with respect to the
basic philosophy, to the use of hard thermal loops (HTL) as
well as to most of notations.

    \vfill
          \fg1                                 

Irrespective of $\vc q \neq \vc 0$ there are the three possible
origins of $O(g)$ terms as specified by Braaten and Pisarski
in \S ~4.3 of \cite{BP}. The 'third' subset (1-loop soft) is
determined by fig.~1 and will be seen in \S ~4 to form a
separate gauge-fixing independent set. The 'second' subset
(1-loop-hard minus leading) can be shown to be less than
$O(g)$ in much the same way as in \I. Most probably, the
'first' subset (2-loop hard) remains below $O(g)$, too.
Admittedly, we did not study these 2-loop diagrams at
$q \neq 0$. Instead we trust in the argument (given at the end
of \S ~3 of \I) that $O(g)$ can only be reached when the scale
$m^2 = g^2 N T^2 / 9$ is involved in the loop integrals.
We also refrained from numerical work. Rather, we focus on
properties and structure. But note that if the plasmon
frequency $\o = m f( g, q/m )$ was known quantitatively one
would be able to look at the line $\o =0$ in the $g$-$q$-plane
and to learn about the phase transition. Lowering the
temperature, the lowest value of $g$ is reached first.

After collecting known details on the (true) leading order in
\S ~2, we shall test the matrix $\d\P\omn$, as determined by
fig.~1, by comparing its properties such as transversality with
known general predictions (\S ~3). These are exploited in \S ~4
for the construction of a suitable scheme to calculate
$\d\P_\ell$. Several properties of the algebraic result are then
deduced as listed in the abstract. The behaviour of $\d\P_\ell$
near the lightcone is studied separately in \S ~5, followed by a
short conclusion in \S ~6.
%
%
\let\dq=\thq \renewcommand{\theequation}{2.\dq}
\setcounter{equation}{0}

\parag {2. \ Notations and identities }

To work with the Lagrangian \eq{1Lag} in thermal field theory
\cite{Kapu,LW,GPY} we introduce the number of gluons ($N^2-1$),
the time contour (Matsubara), the metric (\hbox{$+ - - -$}), the
Bose function ($\,n(p)= 1/(e^{\beta p} -1)\,$) and a hard-soft
threshold ($q^\ast$). Let $Q = (i\o_n , \vc q )$ be the external
momentum running through fig.~1, and $P$ the loop momentum.
Throughout {\sl this} paper, the capital $K$ has the fixed
meaning $K \equiv Q-P$, and, correspondingly, $\vc k \equiv
\vc q - \vc p$. If $\D (P)$ is any function of $P$, then $\D^{-}$
stands for $\D (K)$. The thermal propagator at $O(1)$ reads
\be  \label{2G}
   G\omn (P) = \Cc\omn (P) + \alpha \D_0^2(P) P^\mu P^\nu
   \;\; \mbox{with} \;\;
   \Cc\omn (P) \equiv \D_t (P) A\omn (P) + \D_\ell (P) B\omn (P)
   \;\; ,
\ee
where $\D_0 = 1/P^2$ and
$\D_{\ell\, ,\, t} = 1 / ( P^2 - \P_{\ell\, ,\, t } (P) \, ) \,$.
The Lorentz matrices in \eq{2G} are members of the basis
\bea  \label{A-D}
   & & \hspace{-1cm}
   A= g-B-D \;\; , \;\; B= { V \circ V \over  V^2 } \;\; , \;\;
   C= { P \circ V + V \circ P  \over \wu 2 P^2 p } \;\; , \;\;
   D= { P \circ P \over  P^2 } \qquad \\
\label{V}  & & \hspace{-1cm}
   \mbox{with}  \quad V= P^2 U - (U \cdot P) \, P = ( - p^2 \, ,
   \, - P_0 \vc p \, )  \qquad
\eea
and $U=(1 \, , \vc 0 \, )$ the four-velocity of the thermal
bath at rest. The functions $\P_{\ell\, ,\, t} (P)$, related
by $\P_\ell + 2 \P_t = 3 m^2$, derive through $\Tr (B \P )$
and ${1\02}\Tr (A \P )$, respectively, from the $O(1)$
polarization tensor $\P\omn$ \cite{FT,BPward}
\be  \label{2YY}
  \P\omn = 3m^2 \( \, U^\mu U^\nu - (UP) \int_\O
  { Y^\mu Y^\nu \0 YP } \; \)
\ee
where $Y \equiv (1,\vc e )$ with $\vc e$ a unit vector. The
angular integral $\int_\O$ over the directions of $\vc e$ is
normalized to one$\,$: $\int_\O Y = U\,$. To obtain the 3--leg
and 4--leg HTL vertices in this
nice ''$Y$-language'' (cf. \eq{4MWX} to \eq{4HIJ} below) one
conveniently starts off from the effective action \cite{eff}.

Besides $\Cc\omn$ in \eq{2G} there is another useful tensor
\be   \label{2p}
   \Pc\omn (P) \; \equiv \; P^\mu P^\nu - P^2 g\omn
   + \P\omn (P)  \; = \; \( \P_t -P^2 \) A\omn
   + \( \P_\ell -P^2 \) B\omn  \;\; .
\ee
Both, $\Cc$ and $\Pc$, are symmetric in the Lorentz indices and
are projectors in the sense $\Cc (P) \cdot P = 0$. The product
of $\Cc$ with $\Pc$, used repeatedly in the sequel, has the neat
property
\be  \label{2chip}
  - \Cc^{\mu\rho} (P) \Pc_{\rho\nu} (P)
  \; = \; g^\mu_{\;\;\nu} - D(P)^\mu_{\;\;\nu} \;\; .
\ee
In deriving \eq{2chip}, the properties $A^2=A$, $B^2=B$, $AB=0$
have been exploited.

The following 'perturbative' Ward identities of hot thermal
QCD are usually established by direct verification. However,
they can also be derived \cite{Schr} from BRS invariance of the
effective action$\,$:
\be  \label{2W3}
  (Q_3)_3 \fG^{123} (Q_1 , Q_2 , Q_3 )
  = \Pc^{12} (Q_1 ) - \Pc^{12} (Q_2 ) \;\; ,
\ee
\be \label{2W4}
  (Q_4)_4 \fG^{1234} (Q_1 , Q_2 , Q_3 , Q_4 )
  =  \fG^{123} (Q_1 , Q_2 + Q_4 , Q_3 )
     - \fG^{123} (Q_1 + Q_4 , Q_2 , Q_3 ) \;\; .
\ee
Here, the small numbers refer to Lorentz indices, while
summations over colours are already done \cite{BP}. Momentum
conservation $\sum Q_i = 0$ is understood. $\fG$ is
''$\bigodot$'' in fig.~1, i.e. the sum of the bare (3-- or 4--)
vertex and the HTL part$\,$: $\fG = \bG + \G$, and $\G$ is the
HTL. We avoid the common notation ''$\d\G$'' to emphasize that
the two elements of $\bigodot$ are equal-rank partners. The
symmetry properties of $\fG$ are
\be  \label{2sy3}
  \fG^{1,2,3}(Q_1,Q_2,Q_3) = \fG^{312}(Q_3,Q_1,Q_2)
  = -\fG^{132}(Q_1,Q_3,Q_2) = -\fG(-Q_i) \;\; ,
\ee
\be \label{2sy4}
    \fG^{1234}(Q_1,Q_2,Q_3,Q_4) = \fG^{3412}(Q_3,Q_4,Q_1,Q_2)
  = \fG^{2143}(Q_1,Q_2,Q_3,Q_4) \;\; .
\ee

All the relations so far listed are details of the true zeroth
order. But the difference
\be   \label{2dp}   {} \hspace{-.3cm}
  \d\P\omn = {g^2 N \0 2} \sum \( G_{\lambda \rho}
  \fG^{\mu \nu \lambda \rho} + G_{\rho \sigma}^{-}
    G_{\tau \lambda} \fG^{\mu \sigma \tau} \,
  \fG^{\nu \rho \lambda}  - \D_0^{-} \D_0 \lk 10 P^\mu P^\nu
  - 4 P^2 g\omn \rk  \) \;\; , \quad
\ee
is a true first order object (see (4.2) of \I). The three terms in
the round bracket still correspond to the three elements of
fig.~1. The arguments of the 4--$\!\fG$ are $Q,-Q,-P,P\,$, those
of both 3--$\!\fG$'s are $Q,-K,-P\,$. The blank summation symbol
means
\be   \label{2sum}
  \sum \equiv \sum^{\rm soft}_P
  = \( 1 \0 2\pi \)^3 \int \! d^3p \; \sum_{P_0} \qquad , \qquad
  \sum_{P_0} = T\sum_n \;\; .
\ee
%
%
\let\dq=\thq \renewcommand{\theequation}{3.\dq}
\setcounter{equation}{0}

\parag {3. \ Transversality }

\noindent
3.1 KNOWN EXACT PROPERTIES OF $\P$

As is well known, the BRS-invariance of the gauge-fixed
Lagrangian \eq{1Lag} may be exploited to derive some exact
relations. The one for the (full $=$ overlined) two-point
function $\ov{G}$ reads $\,Q^\mu \, \ov{G}\mn (Q) \, Q^\nu =
\alpha\,$ \cite{BBJ} and can be extended to other gauges
\cite{KoKuMa,Ku}. This relation, if rewritten with the expansion
$\ov{G} = \ov{\D}_t A +\ov{\D}_\ell B +\ov{\D}_c C +\ov{\D}_d D$,
turns into the first equation in \eq{3ex1} below. On the other
hand, one may use Dysons equation
$\ov{G}=G_0 + G_0 \ov{\P}\, \ov{G}$, plug in the expansions of
$G_0$, $\ov{G}$ and $\ov{\P}$ and derive the equation to
the right$\,$:
\be    \label{3ex1}
  \ov{\D}_d = {\a \0 Q^2} \qquad , \qquad
  \ov{\D}_d = { \alpha \( Q^2 - \ov{\P}_\ell \) \0
   \( Q^2 - \alpha \ov{\P}_d \) \( Q^2 - \ov{\P}_\ell \)
   + {1\02}\alpha \ov{\P}_c^2 }   \;\; .
\ee
Equating the two versions in \eq{3ex1}, one arrives at the
following exact relation \cite{KoKuMa}, made explicit and
discussed by Kunstatter \cite{Ku}$\,$:
\be  \label{3ex2}
  2 \( Q^2 - \ov{\P}_\ell \) \ov{\P}_d = \ov{\P}_c^2  \;\; .
\ee
We exclude a hypothetical term $\sim 1/( Q^2 - \ov{\P}_\ell )$
in $\ov{\P}_d$, since it would make the propagator $\ov{\D}_\ell
= (Q^2 - \a \ov{\P}_d ) / [ Q^2 ( Q^2 - \ov{\P}_\ell ) ]$
quadratic singular. Then, from \eq{3ex2}, the exact coefficient
$\ov{\P}_c$ vanishes on the longitudinal mass-shell. Hence, all
terms of the small-$g$ asymptotics of $\ov{\P}_c$ do so as well.
To utilize this fact for the (true) first-order term $\d\P$, we
insert $\ov{\P}_i = \P_i + \d\P_i + ...$ into \eq{3ex2}, remember
that $\P_d=0$ and $\P_c=0$, read \eq{3ex2} at order $O(g)$ to get
$( Q^2 - \P_\ell )\, \d\P_d =0$, i.e. $\d\P_d=0$  (since $\d\P$
cannot develop a delta-function perturbatively), notice that
$\d\P = \d\P_t A + \d\P_\ell B + \d\P_c C$ and infer that
\be  \label{3trans}
  Q^\mu \d\P\mn (Q) \; = \; 0 \qquad \mbox{ if } \qquad
  Q^2 - \P_\ell (Q) = 0 \;\; .
\ee
Thus, $Q \d\P$ vanishes on the longitudinal mass-shell.
Whatever a detailed calculation of $\d\P\omn$ brings about,
it must respect the restricted transversality \eq{3trans}.

\newpage
\noindent
3.2 TESTING $\,\d\P$

We shall study $Q \d\P$, with $\d\P$ given by \eq{2dp},
to see how the longitudinal mass-shell condition makes this
expression vanish. We write
$Q^\mu\d\P\mn = Q\d\P_{\; \nu}^{{\rm tadpole}}
 + Q\d\P_{\; \nu}^{{\rm loop}} - Q\d\P_{\; \nu}^{[-]} $.
The last term can be simplified as
\be  \label{3Rmin}
  Q\d\P_{\; \nu}^{[-]} = g^2N \sum\D_0 (K)\, Q^\sigma
  \( g_{\sigma\nu} - D_{\sigma\nu}(P) \) \;\; .
\ee
For the tadpole contribution the identities \eq{2W4}, \eq{2sy3}
lead to
\be  \label{3RT2}
  Q\d\P_{\; \nu}^{{\rm tadpole}} = g^2 N \sum
  G(P)^{\rho \sigma}\fG_{\nu\sigma\rho} (Q,-K,-P) \;\; ,
\ee
where in a last step the sign of $P$ has been reversed (leaving
$G$ invariant) in one of two terms. Remember that $K = Q-P$.
In the loop contribution,
\be
  Q\d\P_{\; \nu}^{{\rm loop}} = { g^2 N \0 2 } \sum
  G(K)^{\rho \sigma} Q^\mu \fG_{\mu \sigma \tau}(Q,-K,-P)\,
  G(P)^{\tau \lambda\,} \fG_{\nu \rho \lambda} (Q,-K,-P) \;\; ,
\ee
the product $Q \fG$ may be replaced by $\Pc(K) - \Pc (P)\,$.
Since the remaining three factors $GG\fG$ are odd under the shift
$P \to K\,$, this $\Pc$-difference my be replaced by
twice of one term. Hence, using \eq{2chip},
$( Q \fG G )_\sigma^{\;\;\lambda}$ may be replaced by
$ ( - \Pc G )_\sigma^{\;\;\lambda}
  = g_\sigma^{\;\;\lambda} - D(P)_\sigma^{\;\;\lambda}\,$.
Obviously, the $g$-term cancels when the tadpole- and
loop-results are added. The factor $P^\lambda$ in the $D$-term
allows to use the identity \eq{2W3} once more:
\be  \label{3RTL}
  Q\d\P_{\; \nu}^{{\rm tadpole + loop}} = g^2 N \,
  \sum \D_0(P) G(K)^{\rho \sigma} P_\sigma
  \( \Pc_{\nu \rho} (Q) - \Pc_{\nu \rho} (K) \)  \;\; .
\ee
For the term containing $\Pc(K)$ we remember \eq{2chip}, i.e.
$G(K) \cdot \Pc (K) = D(K) - g$. By a shift $P \to K$ we
see that it equals the subtraction term \eq{3Rmin}. The term in
\eq{3RTL} containing $\Pc(Q)$ may be written as
\be  \label{3pvQ}
   g^2 N \Pc(Q)_{\nu \rho} \sum \D_0(K)
   \( \Cc(P)^{\rho \sigma} Q_\sigma + \alpha \D_0^2(P) (PQ)
   P^\rho - \alpha \D_0(P) P^\rho \) \;\; .
\ee
Symmetrizing the third term of \eq{3pvQ} with respect to
$P \to K$ it becomes proportional to $Q^\rho$ and drops out
via $\Pc \cdot Q = 0$. Thus,
\be  \label{3RRR}
  Q^\mu \d\P_{\mu \nu} (Q) = g^2 N \Pc_{\nu \rho}(Q) \;
  \zeta^\rho (Q) \quad \mbox{with} \quad \zeta^\rho (Q)
  = \sum \D_0(K) G(P)^{\rho \sigma} Q_\sigma \;\; .
\ee
The sum $\zeta^\rho$ ''knows'' of only $Q_0$ and $\vc q$.
Therefore, it must be a linear combination of $Q^\rho$ and
$U^\rho$, or, equivalently, of $Q^\rho$ and $V^\rho\,$: \
$\zeta^\rho (Q) = c_q Q^\rho + c_v V^\rho \,$. Using this
in \eq{3RRR}, and $\Pc \cdot Q = 0$, $Q\cdot V = 0$ as well as
$\P_{\nu \rho} V^\rho = \P_\ell V_\nu$, we end up with
\be  \label{3end}
  Q^\mu \d\P\mn (Q) = V_\nu \, g^2 \, N \( \P_\ell (Q) - Q^2 \)
  \, c_v \;\; .
\ee
\eq{3end} is the desired result. The coefficient
$c_v = V_\rho \sum \D_0 (K) G(P)^{\rho \sigma} Q_\sigma / V^2$ is
non-zero. Thus, $\d\P\omn\,$ is transverse only on the longitudinal
mass-shell. This agreement with \hbox{Kunstatter} \cite{Ku}
confirms our set-up of contributions to $\d\P\omn\,$.

   There is a bit more of information in \eq{3end}. Multiplication
with $V^\nu$ gives $\d\P_c = g^2 N\wu 2 q$ $( Q^2 - \P_\ell )
c_v $. But completing the sandwich by $Q^\nu$ we get $\d\P_d = 0$
(all $Q$). Hence, $\P_d$ starts with a term below $O(g)$, which
we call $\d^2\P_d$. Now, equating the first non-vanishing terms
on both sides of the exact relation \eq{3ex2} and cancelling
a factor $( Q^2 - \P_\ell )$ on both sides, we obtain
\be  \label{3below}
 \d^2 \P_d = \( g^2 N q \, c_v \)^2 \( Q^2 - \P_\ell\)  \;\; .
\ee
Thus, even the less-than-$O(g)$ term in $\P_d$ vanishes on the
longitudinal mass-shell.
%
%
\let\dq=\thq \renewcommand{\theequation}{4.\dq}
\setcounter{equation}{0}

\parag {4. \ The longitudinal next-to-leading order term
           $\d\P_\ell$  }

To evaluate the sandwich $\d\P_\ell = (V \d\P V)/V^2$ on the
longitudinal mass-shell, the restricted transversality
\eq{3end} may be exploited to simplify each of the sandwiching
vectors as $V = Q^2 U - (QU) Q \;\to \; Q^2 U$:
\be  \label{400}
   \d\P_\ell = - {Q^2 \0 q^2} \d\P_{00} \quad \mbox{at}
   \quad Q^2 = \P_\ell(Q)  \;\; ,  \;\;
\ee
with $\d\P_{00}$ to be read off from \eq{2dp}.

There is no gauge-fixing dependence of $\d\P_{00}$ on the
longitudinal mass-shell. To check this, consider for example
the $\alpha^2$ term. It derives from the loop contribution
(at $\mu=\nu=0$), with $G$ reduced to $\alpha \D_0^2(P) P^\mu
P^\nu \equiv f(P) P^\mu P^\nu$. Using the Ward identity \eq{2W3}
twice, and $P^\tau \Pc_{\mu \tau} (P) = 0 $, one arrives at
\be  \label{4ff}
  \d\P_{00} \vert _{f^2} = \Pc_{0\rho}(Q) \Pc_{0\tau} (Q)
  {g^2 N \0 2} \sum f(K) f(P) P^\rho P^\tau \;\; .
\ee
But, with view to \eq{2p}, the matrix $A$ has no 0-elements,
hence $\Pc_{0\tau} = \( \P_\ell - Q^2 \) B_{0\tau}(Q)$, and
\eq{4ff} vanishes on the mass-shell. Note that this procedure
working is much more convenient than that in \I. In a similar
manner one can verify that the term linear in $f$ cancels that
of the tadpole (the subtraction term has no $f$). Thus,
$\d\P_{00}$ is independent of $\alpha$. Moreover, it is
an {\sl invariant} under changes of the (even) function $f(P)$.

   We shall exploit this invariance by the following
special choice,
\be   \label{4rep1}
   f(P) = \D_0 (P) \D_{\ell t} (P) {P^2\0p^2} + \D_0 (P)
          \D_\ell (P) \quad \;\; \( \, \D_{\ell t}
          \equiv \D_\ell - \D_t \; \) \;\; , \;
\ee
which leads, via $A=g-B-D$ and
$-P^2p^2 B(P) = (P^2 U - P_0P) \circ (P^2 U - P_0 P)\,$,
to the replacement
\be   \label{4rep2}
  G\omn \;\; \to \quad \D_t \, g\omn
  - \D_{\ell t} {P^2\0p^2} U^\mu U^\nu + \D_{\ell t} {P_0 \0 p^2}
  \lk U^\mu P^\nu + P^\mu U^\nu \rk \;\; . \;
\ee

The four-vectors $P$ in \eq{4rep2} multiply $\fG$'s and reduce
them through the Ward identities \eq{2W3}, \eq{2W4}. But with
the $U$-vectors, we force more Lorentz indices to zero. The
corresponding analysis is straightforward (somewhat lengthy, but
not tedious) and leads to the following ''algebraically final
result''$\,$:
\be  \label{4afin}
  \d\P_\ell = \( 1 - {\omega^2 \0 q^2} \) \, g^2 N \;
  \sum \( c_0  +  \D_\ell^- \D_\ell^{} \, c_{\ell\ell}
          + \D_\ell^- \D_t^{} \, c_{\ell t}
          + \D_t^- \D_t^{} \, c_{tt} \)  \qquad
\ee
with the ''coefficients'' $c$ given by
\bea    \label{4c0}
 c_0 &=& - 2 \D_0 - 4 \D_0^- \D_0 p^2  \\
 \label{4cll}
 c_{\ell\ell} &=& {P^2 K^2 \0 2p^2 k^2 } \X^2  + {1\04} \W
  \( {K^2 \0 k^2 } \d_\ell^{} - {P^2 \0 p^2} \d_\ell^- \)  \\
 \label{4clt}
 c_{\ell t} &=& - {P^2 K^2 \0 p^2 k^2 } \X^2  -
    {K^2 \0 k^2} \fY^2 +  \d_\ell^- {P^2 \0 2p^2} \W \\
 \label{4ctt}
 c_{tt} &=& {P^2 K^2 \0 2p^2 k^2 } \X^2
    + {K^2 \0 2k^2} \fY^2
    + {P^2 \0 2p^2} \fY_-^2 + {1\02} \fZ^2
    + { P_0 K_0 \0 P^2 K^2} \,\d_\ell^- \d_\ell^{}
    - {3 \d_t \0 2} - {3 \d_t^- \0 2}  \;\; . \quad
\eea

   The terms in \eq{4afin} are grouped such that they have neat
properties at small $q$ (see point {\bf (iii)} below). \eq{4c0}
includes the subtraction term $\d\P^{[-]}$. The objects $\d$,
which originate from $P^2 \Pc_{00} = p^2 \d_\ell$, are inverse
propagators,
\be  \label{4delta}
   \d_t^{} \equiv P^2 - \P_t^{} (P) \quad , \quad \d_\ell^{}
   \equiv P^2 - \P_\ell^{} (P) = 3P^2 -3m^2 - 2\d_t^{} \;\; .
\ee
Clearly, \eq{4afin} may be rewritten immediately by (further)
cancellations $\D \d = 1$. As before, an index minus refers
to the substitution $P \to Q-P \equiv K$. The (full)
vertex functions are hidden in the capital letters
\bea  \label{4WX}
 \W = \G_{0000} (Q,-Q,-P,P) \;\; &,& \;\;
 \X = \G_{000} (Q,-K,-P) \quad ,  \\
 \label{4YZ} \hspace{-.4cm}
 \fY^2 = \fG_{00}^{\;\;\;\;\mu} (Q,-K,-P)
       \,\fG_{00\mu} (Q,-K,-P) \;\; &,& \;\;
 \fZ^2 = \fG_0^{\;\;\mu\nu} (\ldots )
       \,\fG_{0\mu\nu} (\ldots ) \quad , \quad \;\;
\eea
There
is no bare part in $\W$ and $\X$ (therefore no star). To
separate the HTL-pieces, $\Y^2 =\G_{00}^{\;\;\;\;\mu} \G_{00\mu}$
and $\Z^2 =\G_0^{\;\;\mu\nu} \G_{0\mu\nu}$, in $\fY^2$ and
$\fZ^2$, respectively, one is led into further analysis with
further use of the Ward identity \eq{2W3}$\,$:
\bea  \label{4schlY}
  \fY^2 &=& \Y^2 + \( 2P_0-4Q_0 \) \X
  - 4 {p^2\0P^2} \,\d_\ell^{}  + 2 {k^2\0K^2}\,\d_\ell^-
  + 5p^2 -4k^2 -4q^2   \\    \label{4schlZ}
  \fZ^2 &=& \Z^2
  - 2 {p^2\0P^2} \,\d_\ell^{}  - 2 {k^2\0K^2}\,\d_\ell^-
  + 3 (2P_0 - Q_0)^2 + p^2 + k^2 -8q^2 \;\; . \quad
\eea
$\W$ and $\X$ change sign under $P \to K$. $\,\Y^2$ and
$\Z^2$ and $\fZ^2$ are invariants under this transformation,
but $\fY^2$ is not.

   Using the $Y$-language (cf. \eq{2YY}), the above HTL-parts
can be dealt with as
\bea  \label{4MWX}
 \W = - 6m^2 P_0 \6_{Q_0} H - \dit \quad
     &,& \quad \X = 3m^2 P_0 H - \dit \quad , \\
 \label{4MY}
 \Y^2 = \X^2 - \vc y ^2  \quad
     &,& \quad \vc y = 3 m^2 P_0 \, \vc I - \dit \quad , \\
 \label{4MZ}
 \Z^2 = \X^{\bf 2} - 2 \vc y ^2 + \Tr \( \zzz \zzz \) \quad
     &,& \quad \zzz = 3m^2 P_0 \, \JJJ - \dit \quad
\eea
with the following angular integrals involved
\be  \label{4HIJ}
        H = \int_\O {1 \0 (YQ) (YP) } \quad , \quad
    \vc I = \int_\O {\vc e \0 (YQ) (YP) } \quad , \quad
  \JJJ = \int_\O {\vc e \circ \vc e \0 (YQ) (YP) } \;\; .
\ee
Compared to \I, the whole trouble of the present non-zero wave
vector analysis stems from these three integrals, see also
\eq{5feyn} below. In order to do the summations over $P_0$, the
angular integrations in \eq{4afin} may be simply shifted to the
left and commuted with $\sum\,$.

So far, the result \eq{4afin} is only recognized to be a
pretty lengthy expression. Next, we enumerate some of its
general properties and limiting cases.

\vspace{.1cm} \noindent
\pt i
{\bf UV-convergence}.
The expression \eq{4afin} is restricted to soft-momentum
contributions automatically, i.e. it does not depend on the
cutoff $q^*$, which bounds the soft scale from above.
For an immediate check, one may reduce the dressed propagators
to bare ones and omit the HTL-parts $\W$ to $\Z^2$. By retaining
only UV-dangerous terms (those occurring in $c_0$), \eq{4afin}
becomes $\sum \( c_0 - c_0 \)$, as expected. In passing, the sum
over $P_0$ in front of \eq{4afin} is convergent, of course.
There is danger only in the terms containing $\W \sim 1/P_0$.
Adding them together gives a finite difference.

\vspace{.1cm} \noindent
\pt {ii}
{\bf IR-convergence}.
The real part of $\d\P_\ell$ is finite in the infra-red along
the whole real $q$-axis. However, there are mass-shell
singularities in the imaginary part for real $q$ as well as
in the real part for imaginary $q$, which are investigated in
a forthcoming paper with A. K. Rebhan \cite{frosch}.
See also point (iv).

\vspace{.1cm} \noindent
\pt {iii}
{\bf Check at $q=0$}.
    As we know from \I, $\d\P_\ell$ is regular when
$q\to 0\,$. Thus, the singular prefactor $q^{-2}$
in \eq{4afin} must be compensated by a vanishing sum
$\sum (c_0 + \ldots )$. Indeed, its expansion in powers of $q$
starts with a $q^2$ term. To show this, we consider the sum
over $c_0$ first. It can be evaluated at arbitrary $q\,$ as
\be \label{4sumc0}
  \sum c_0 = - {4q^2 \0 3m^2} \sum_P^{\rm soft} \D_0(P)
           = q^2 {2 T \0 3m^2 \pi^2 } \int_0^{q^*} dp \;\; .
\ee
Here, a term less than $O(g)$ had been neglected (see (5.35)
of \I), and the mass-shell condition was used. For the last step
in \eq{4sumc0} see (5.9) of \I. The prefactor $q^2$ in
\eq{4sumc0} is to our likings. The cut-off $q^*$ just tells us
that there are also other $q^2$ contributions which compensate
for it. For the other three sums in \eq{4afin}, we need the
integrals \eq{4HIJ} at $q=0$, i.e. $YQ=Q_0\,$:
\bea
   H_0 &=& {1\0 Q_0P_0} \( 1 + \V \) \quad , \quad
   \JJJ_0 = {1 \0 2 Q_0 P_0} \( 1 - {P^2\0p^2} \V
   + {\vc p \circ \vc p \0 p^2} \lk 3 {P^2\0p^2} \V
       + 2 \V -1 \rk  \) \quad ,  \nonu \\
   \label{4H0}
   \vc I _0 &=& {1\0Q_0} \, { \vc p \0 p^2 } \V
   \qquad\quad \mbox{with} \quad \V \equiv {p^2\0P^2}
   {1\03m^2} \P_\ell (P) \;\; .
\eea
{}From \eq{4H0}, the limiting values of $\W$, $\X$, $\fY^2$
and $\fZ^2$ may be obtained (keep bare and HTL-parts
together!). Through shifts $P\to K$ and
$P_0 \to -P_0$ under the sum, one finally obtains
\be  \label{4q0}
 \sum \D_\ell^- \D_\ell^{} c_{\ell\ell} \,\to \, 0
 \;\; \mbox{and} \;\;
   - \sum  \D_\ell^- \D_t^{} c_{\ell t} \;\; , \;\;
   \sum  \D_t^- \D_t^{} c_{t t} \to
   {1 \0 Q_0^2} \sum \D_t \( \d_\ell^{} - \d_\ell^- \) \;\;
\ee
which exhibits the desired compensation.

When the whole expression \eq{4afin} is regarded in the limit
$q\to 0$, it must turn into the final result of \I, which
is eq. (6.1) there. To verify this, the integrals \eq{4HIJ} were
to be expanded including $q^2$. For the simplest term, which is
$c_{\ell\ell}$ in \eq{4afin}, one arrives straightforwardly
at ${\cal M}_3$ in (6.2) of \I. But to handle the immense total
number of terms, we had several sessions with Miss MAPLE
\cite{maple}. She produced the desired answer with all details.

\vspace{.1cm} \noindent
\pt {iv}
{\bf Check at $\o=0$}.
If we leave the real $q$-half-axis and go to purely imaginary $q$,
the longitudinal mass-shell condition $Q^2=\P_\ell (Q)$ may be
followed down to zero-frequency. There, $\d\P_\ell$ (real part)
determines the Debye screening length \cite{Reb}. Along this
line, $\d\P_\ell$ always stays an UV-convergent expression
(see {\bf (i)}), while its IR-singularity is hidden in \eq{4afin}
with \eq{4clt}, see \cite{frosch}. With $Q_0 \to 0$,
all terms in \eq{4afin} which contain HTL-vertices like $\X$
vanish. To derive this we performed the frequency sums first and
the analytical continuation $Q_0 \to \o + i \e
\to 0$ afterwards. It is then a rather easy task to do the
remaining sums (along the lines given in \I). The result is
\be   \label{4o0}
  \d\P_\ell (0,q) = g^2 N T \( 1\0 2\pi \)^3 \! \int \! d^3p
  \( {1 \0 3m^2 + p^2 } - {1 \0 p^2}
  + { 2 \( 3 m^2 - q^2 \) \0 p^2 \( 3 m^2 + k^2 \) } \)
\ee
and agrees essentially with eq. (13) of Rebhan \cite{Reb}
when restricted to the longitudinal mass-shell on which we
stayed from the outset and which is at $q^2=-3m^2$, $\o=0\,$.
In \cite{Reb} this result was obtained in the simplified
resummation scheme of Arnold and Espinosa \cite{AE}, which
differs in that the second term in \eq{4o0} comes with a
different sign, making the whole expression UV divergent.
With dimensional regularization, which is required in the
method of \cite{AE}, this leads to exactly the same result
as \eq{4o0} whose manifest UV finiteness comes from
the explicit subtraction of the one-loop bare piece in
\eq{2dp} \footnote{A. K. Rebhan, private communication}.
Note that $m^2$ of Rebhan is $3m^2$ in our notation.

\newpage
%
%
\let\dq=\thq \renewcommand{\theequation}{5.\dq}
\setcounter{equation}{0}

\parag {5. \ Approaching the lightcone  }

With increasing (real) wave vector $q$, the plasmon spectrum
$\o (q)$ approaches the lightcone. Hence, there is a small
parameter $\e^2 \equiv \( \o^2 - q^2 \) / q^2$. In this
section we discuss the behaviour of the next-to-leading
order term $\d\P_\ell$ as a function of $\e$ for $\e\to0\,$.
We expect order $1/\e$-contributions to $\d\P_{00} =
- \d\P_\ell \, q^2/Q^2$ from an earlier study \cite{KRS} of
scalar electrodynamics as a toy-model of the gluon plasma.
As in the scalar theory, it turns out that these singularities
signal the need of further resummations beyond the scheme
of Braaten and Pisarski and that $\o (q)$ intersects the
lightcone at some finite value $q^{\rm crit}$ as given in
\cite{KRS} at $O(1)\,$. Here we shall {\sl not} go up to
this point. Instead, we remain interested in properties
of $\d\P_\ell$ only, stay within the methodology of Braaten
and Pisarski, and, thus, defer the construction of a (new)
consistent perturbative scheme for possible future work.

We shall show here, that there are (at least) two origins of
$1/\e$. One is the same as in the scalar theory (index SED).
The other is in the HTL-vertex pieces that remained in
$\d\P_\ell$, \eq{4afin} (remember that parts of the HTLs have
been converted by Ward identities and annihilated through the
mass-shell condition or by cancellations $\D\d=1$). Let us
split \eq{4afin} into the parts just mentioned$\,$:
\be  \label{5s+s}
  \d\P_{00} = - {q^2 \0 Q^2} \d\P_\ell \equiv g^2 N \sum
  \quad \mbox{with} \quad \sum = \sum \( c_0 + \ldots \)
  = \sum\nolimits^{\rm SED} + \sum\nolimits^{\rm HTL} \quad .
\ee
For the SED-part, put $\W = \X = \Y = \Z = 0$ in \eq{4afin}.
$\sum^{\rm HTL}$ is simply the rest.

The two parts form separate UV-convergent sets. If, with view
to the lightcone, the replacements $Q_0^2\to q^2$,
$\P_\ell \to 0$ and $\P_t \to 3m^2/2 \equiv
\mu$ are consequently performed in $\sum^{\rm SED}$, which
implies $\d_\ell$, $\D_\ell$, $P^2\D_t \;\to
 \; P^2$, $\D_0$, $1+\mu^2\D_t$, respectively,
then the sum turns into
\be   \label{5sed}
  \sum\nolimits^{\rm SED} \to \sum \( 2 \lk \D
  - \D_0 \rk + 4 \lk \D^{-} \D - \D_0^{-} \D_0 \rk
  + 4 \mu^2 \D^{-} \D \)  \quad
\ee
with $\D = 1/(P^2-\mu^2)$. In the course of this, terms
$\sum \D^{-} \D$ have been neglected since they contain no $1/\e$
(see eqs. (6.5), (6.10) in \cite{KRS}). The last term in
\eq{5sed}, which is of this type, is included only for better
identification of the above result with eqs. (3.3), (3.8) in
\cite{KRS}. Clearly, for the extraction of $1/\e$ from \eq{5sed}
we may simply refer to \cite{KRS}.

The above argument was based on ''replacements''. Its
justification, however, runs into (non-abelian) difficulties.
Note that inner momenta $P$ were taken at the lightcone, but
actually only the outer $Q$ is placed there. For a rough
argument, consider the formula
\bea  \label{5rot}
 & &  \sum \D_\ell^{-} \D_t f(\vc p ) =   \nonu  \\
 & &  T \( {1\02\pi} \)^3 \!\int\! d^3p \, f(\vc p ) \int\! dx \,
   {1\0 x} \rho_t (x,p) {1\0 x-Q_0} \lk Q_0 \D_\ell (Q_0 -x,k)
   - x \D_\ell (0,k) \rk \;\; , \qquad \qquad
\eea
which is valid if by virtue of $f(\vc p )$ the $p$-momenta are
restricted to be soft. \eq{5rot} generalizes (6.5) of \I to
non-zero $\vc q$. The square bracket becomes large at $x=Q_0
\pm k$, or, if $Q^2=0$, at $\vc p = x \vc q /q $. In this region
of the $x$-$p$-plane, at larger (but still soft) $p$,
$\rho_t$ is dominated by its pole-contribution $ r_t (p) \lk \,
\d (x-\o_t (p) ) - \d (x+\o_t (p) ) \,\rk $ (see Appendix B of \I),
which indeed gives the transversal propagator the ''scalar'' form
$\D_t = c / [ P^2 - \P_t (\o_t(p) , p ) ]$ with $c=2r_t(p)\o_t(p)$.
We add the general definition of a spectral density$\,$:
$ \D (P) = \int \! dx \, \rho (x,\vc p ) \, / \, ( P_0 - x )\,$.

The idea that even the HTL-part in \eq{5s+s} could diverge
at the lightcone comes to mind if the above rough argument
($Q$ at the cone enforces $P$ to be there too) is applied
to the hard loop-integrations inside a HTL-vertex, too.
To exhibit the corresponding mechanism consider a typical
but simple term $\Upsilon\,$:
\be   \label{5toy}
 \Upsilon \,\equiv\, \sum f \( p , \vc p \vc q \) \;
  H^2 ( P_0 ) \; = \; T \( {1 \0 2\pi} \)^3 \!\int\!
  d^3p \; f \( p , \vc p \vc q \) \; H^2 (0) \;\; ,
\ee
where $H$ is the integral in \eq{4HIJ} viewed as a function
of $P_0$, and $f$ is any function restricting the integration
to soft $p\,$. $\Upsilon$ occurs in the last term of \eq{5OL}
below. The summation over $P_0$ is performed to the right in
\eq{5toy}, while $Q_0$ stays an imaginary Matsubara frequency.
$H(P_0)$ is real (symmetrize \eq{4HIJ} with respect to
$\vc e \to - \vc e $), and its sign is dominated by the
hard parts $Q_0$, $P_0$. If $Q_0 P_0 \neq 0$, sign$(H)$ =
sign$(Q_0 P_0) \equiv - \eta$. Now, using Feynman
parametrization (and avoiding vanishing denominators),
the angular integration can be done$\,$:
 \be   \label{5feyn}
    {1 \0 a\, b } = \int_0^\infty \! dv \; { 1 \0 (a+vb)^2 }
    \quad \Rightarrow \qquad
    H (P_0) =  \int_0^\infty \! dv \; { \eta \0 Q^2
    + 2 v\,\eta\, PQ + v^2 P^2} \quad . \quad
 \ee
If $P_0 =0\,$, the symmetric version  ${1\02} \lk
\int_{\,{\rm at}\,\eta=+1}\, + \,\int_{\,{\rm at}\,\eta=-1} \rk $
of \eq{5feyn} must be used in accord with a principal value
prescription of \eq{4HIJ} at $P_0=0$.
The result of integrating over $v$,
\be   \label{5anal}
 H(0) = {i \0 2 p \wu {\gamma^2 - Q_0^2} } \,\ln \(
        {   - i p \wu {\gamma^2 - Q_0^2} - \vc p \vc q
         \0 - i p \wu {\gamma^2 - Q_0^2} + \vc p \vc q } \)
        \;\; , \;\; \gamma^2 =  {n \0 p^2 }
        \;\; , \;\; n = p^2 q^2 - ( \vc p \vc q )^2  \;\; ,
\ee
is the right place to do the analytic continuation. In the
complex $Q_0$-plane, there are cuts on the real axis ranging
from $\gamma$ to $\infty$ and $-\gamma$ to $-\infty$. Thus,
through $Q_0 \to \o + i 0$ with $\o \gsim q$,
we arrive at
\be  \label{5log}
    H(0) =  - {1\0 2qp} \, {1 \0 \wu {\e^2 + u^2} } \ln \(
     { \wu { \e^2 + u^2} + u \0 \wu { \e^2 + u^2 } - u } \)
    \quad , \quad u = { \vc p \vc q \0 pq } \;\; .
\ee
Note that with $\e \to +0$, the square of the above
expression becomes $1/\e$ times a representation of the
delta function, by means of which our toy term $\Upsilon$ is
easily evaluated$\,$:
\be  \label{5delta}
   H^2 (0) \;\to \; {1\0\e} \,
           {\pi^3 \0 4 q^2 p^2 } \; \d (u)
   \qquad \Rightarrow \qquad \Upsilon
   = {1\0\e} \, {T \pi \0 16 q^2 } \int_0^\infty
     \! dp \, f(p,0) \quad . \quad
\ee

{}From the above we are led to a handy method for treating
$\sum^{\rm HTL}\,$. Apparently, terms linear in $\X$ etc. can
lead to logarithms of $\e$ only (consider \eq{5feyn} at $Q^2=0$).
Hence we restrict $\sum^{\rm HTL}$ to the terms quadratic in $\X$,
$\Y$, $\Z$. For a further simplification we observe that the
latter two HTLs, $\Y^2$ and $\Z^2$, can be expressed by $\X^2$
\cite{Fle}. For example, the integral $\vc I$ in \eq{4HIJ}
''knows'' of only two spatial directions and is therefore a
linear combination of $\vc q $ and $\vc p\,$, and so on. The
procedure is rather tedious and leads (omitting further terms
linear in $\X$) to the surprisingly simple relations
\be  \label{5xx}
  \Y^2  =  \( 1 - {r\0n} \) \X^2 \quad , \quad
  \Z^2 = 2 \(1 - {r\0n} \)^2 \X^2 \quad
  \mbox{with} \quad
  r = \( P_0 \vc q - Q_0 \vc p \)^2 \;\;
\ee
and $n$ see \eq{5anal}. Note that both, $n$ and $r$ are
invariants under $P\to K$. Using \eq{5xx}, the
HTL-term can be written conveniently as$\,$:
\be   \label{5OL}
   \sum\nolimits^{\rm HTL} = {1\04} \sum \X^2 {1 \0 p^2 k^2}
   \( \O^{-} \O + \Lambda^{-} \Lambda +
   4 q^2 {k^2 \0 n} \Lambda + 4 q^4 {p^2 k^2 \0 n^2} \) \;\; .
\ee
The ''propagators'' in \eq{5OL} were introduced by
$\O = P^2 \( \D_\ell - \D_t \)$ and
$\Lambda = \O - 2 p^2 ( 1 - r / n ) \D_t - 2q^2 p^2 / n\,$.
But we actually need only their spectral densities
\be  \label{5dens}
  \rho_\O = \( x^2 - p^2 \) \( \rho_\ell - \rho_t \)
  \quad , \quad \rho_\Lambda = \rho_\O
      + {2p^2\0n} \( qx - \vc q \vc p \)^2 \rho_t \;\; .
\ee
For $\O$ see the table 1 in \I. It remains to evaluate sums of
three types, $\sum \X^2 \D^{-} \D / p^2 k^2 \,$,
$\sum \X^2 \D / p^2 k^2 $ and $\sum \X^2 / p^2 k^2 $, along the
lines \eq{5toy} to \eq{5delta}. After some analysis \cite{Fle}
and repeated omission of less-divergent terms (less than $1/\e$),
we arrive at
\bea   \label{5fin}
 & &  \sum\nolimits^{\rm HTL} =  - {1\0\e} \, {9 m^4 T \0 32 q^2 }
      \int\! d^3p \; { 1 \0 p^2 k^2 } \, n \, \wu n  \nonu \\
 & &  \lk \, { 1 \0 \vc p \vc q } \rho_t \(
       {\vc p \vc q \0 q} , p \) - { 1\0 \vc p \vc q } \rho_\ell
       \( {\vc p \vc q \0 q} , p \) \, \rk
    \, \lk \, \D_t \( {\vc k \vc q \0 q}, k  \)
       - \D_\ell \( {\vc k \vc q \0 q} , k \) \, \rk \;\; . \qquad
\eea
The expression \eq{5fin} is UV-convergent and IR-finite.
The first square bracket is positive in the whole range
of integration (the densities reduce to their cut-parts, and
$\rho_t^{\rm cut}/\vc p \vc q  \ge 0\,$,
$\rho_\ell^{\rm cut}/\vc p \vc q  \le 0\,$, \cite{HS}).
But the real part of the second square bracket does not have
such a nice property. The propagators contain Landau damping
(see (B.4) and (B.5) of \I), and $\D_\ell$, though dominating,
changes sign. At this point, we abstain from a more detailed
(numerical) analysis. In short, HTL vertices {\sl do} contribute
to the singularity at the lightcone.

\newpage
%
%
\let\dq=\thq \renewcommand{\theequation}{6.\dq}
\setcounter{equation}{0}

\parag {6. \ Conclusions  }

The next-to-leading order calculations on the gluon plasmon
dispersion known so far are extended to arbitrary
wave vectors $\vc q\,$. The real part of the plasmon
self-energy $\d\P_\ell$ (although remaining a lengthy, still
algebraic, expression) is found to have all the expected
properties, such as gauge-fixing independence, convergence in
the UV and IR, and the correct limiting behaviour at
$q \to 0$ as well as at $\o \to 0$ along the
longitudinal mass-shell line. Close to the lightcone, two
mechanisms are detected which violate the common $O(g)$-scheme
in this limit, since they let $\d\P_\ell$ diverge as $1/\e$,
i.~e. stronger than the $\ln (\e )$ in the leading order.
The study of the longitudinal dispersion near the lightcone
needs a new consistency scheme which is still unknown.

%
{\sl We are grateful to Gabor Kunstatter, Anton K. Rebhan and
     York Schr\"oder for valuable discussions. }
%
%
    \vspace{1cm}   \renewcommand{\section}{\paragraph}

\vspace{3cm} \centerline{$\Upsilon$}

\end{document}